# Adaptive coded illumination Fourier ptychography microscopy based on physical neural network

**Ruiqing Sun,**[1] **Delong Yang,**[1] **Yao Hu,**[1] **Qun Hao,** [1, 2, 4] **Xin Li,**[3] **and Shaohui Zhang,** [1, *]

[1]*School of Optics and Photonics, Beijing Institute of Technology, Beijing 100081, China*
[2]*Changchun University of Science and Technology, Changchun 130022, China*
[3]*Department of General Surgery, Xiangya Hospital, Central South University, Changsha 410011, China*
[4]*e-mail: qhao@bit.edu.cn*
[*]*Corresponding author: zhangshaohui@bit.edu.cn*

**Abstract:** Fourier Ptychographic Microscopy (FPM) is a computational technique that achieves a large space-bandwidth product imaging. It addresses the challenge of balancing a large field of view and high resolution by fusing information from multiple images taken with varying illumination angles. Nevertheless, conventional FPM framework always suffers from long acquisition time and a heavy computational burden. In this paper, we propose a novel physical neural network that generates an adaptive illumination mode by incorporating temporally-encoded illumination modes as a distinct layer, aiming to improve the acquisition and calculation efficiency. Both simulations and experiments have been conducted to validate the feasibility and effectiveness of the proposed method. It is worth mentioning that, unlike previous works that obtain the intensity of a multiplexed illumination by post-combination of each sequentially illuminated and obtained low-resolution images, our experimental data is captured directly by turning on multiple LEDs with a coded illumination pattern. Our method has exhibited state-of-the-art performance in terms of both detail fidelity and imaging velocity when assessed through a multitude of evaluative aspects.

## 1. Introduction

With the development of life sciences, microscopic biological structure information has been obtaining increasing interest and attention among the researchers [1-3]. However, due to the Space-Bandwidth Product (SBP) limit of conventional optical microscopes, there is a growing conflict between the field of view (FOV) and spatial resolution, which are both crucial for observing dynamic processes across different spatial and temporal scales [4]. To overcome this limit, Fourier ptychographic microscopy (FPM) has emerged as a typical computational imaging technique that produces high-resolution(HR) images with a wide FOV [5-8]. This technique has attracted considerable attention due to its ability to be extended to different applications, such as digital pathology, drug discovery and stem cell biology [6,9,10]. Traditional large SBP imaging methods always require 2D precise scanning in the spatial domain using a high numerical aperture (NA) objective. However, the scanning process is time-consuming, making it unsuitable for in vitro imaging of dynamic events. In contrast, FPM scans the sample's spectrum by sequentially turning on each LED unit located at different positions on the LED board and combining corresponding low-resolution(LR) images in Fourier space to form a large SBP complex image with a small NA objective [5,6,11,12]. This imaging approach significantly reduces the overall hardware cost. FPM's reconstruction process can be accomplished through the deployment of either nonlinear optimization methodologies or iterative algorithms [5,7,13-15]. As a variety of machine learning techniques have gradually emerged to solve the imaging reconstruction problem [16-23], deep learning-based optimization methods are also used to solve the FPM optimization problem [24,25].

To increase the quality of the reconstructed HR image, it is important to ensure sufficient overlap ratio between adjacent sub-spectrum areas corresponding to adjacent LEDs, with at least 60% overlapping coverage being necessary [26, 27]. In other words, it is necessary to turn on an adequate quantity of LEDs. Consequently, a typical FPM system always faces challenges with long acquisition times due to the large number of raw images and extended exposure times required for dark field images, which may limit its potential applications. There are two main approaches to address this issue: one is to reduce the number of acquisitions for LR images, and the other is to use higher-performance hardware. Generally, employing LED light sources with increased brightness and switching frequency, in conjunction with image sensors boasting higher frame rates, can substantially enhance the overall acquisition efficiency. Nonetheless, hardware modifications often mean more extensive system adjustment cycles, augmented uncertainty, and elevated costs. Therefore, some studies have improved the temporal resolution by turning on multiple LEDs simultaneously to shorten LR images acquiring time [4,14,20]. These studies can be divided into two categories: manual rule-based and data-driven approaches. Manual rule-based methods can provide a universal illumination mode without additional training time and have high interpretability, while data-driven methods can provide a more suitable one. Manually designed modes may lead to satisfactory results for samples with specific spectral distribution types. However, for samples exhibiting significant differences in spectral distribution characteristics, the reconstruction outcomes can sometimes be unsatisfactory. Data-driven methods can perform better in generating specific illumination modes, but the gold standard required for training can be scarce, particularly in some fields such as materials science and medicine. To deal with the aforementioned challenges, this paper proposes an unsupervised physical neural network to create adaptive illumination modes for any specific samples. Our model is established within a highly scalable framework, which is also employed in the Fourier Ptychographic Multi-parameter Neural Network (FPMN) [24]. Each layer within our model possesses a distinct physical interpretation, as the physical priors of FPM are incorporated. It is important to note that previous illumination optimization methods [22, 27] often produce a mode consisting of LEDs with varying brightness levels. The complex brightness ratio between different LEDs heighten increases the hardware construction requirements, and information associated with relatively low brightness is often susceptible to loss during capture. As a result, these methods typically rely on linear combinations of LR images from single LED sequential illumination, rather than directly activating multiple LEDs and taking corresponding composite LR images. In contrast, we turned on LEDs based on our generated mode to capture directly instead of digital combination with acquired raw LR images.

By reviewing the illumination mode generation task, we found that with appropriate prior knowledge, it is possible to generate a specific coded illumination mode for each sample based on its Fourier domain distribution. In the entire imaging process, we update the encoded illumination layer and the sample layer synchronously to generate a specific illumination mode with the best reconstruction quality, based on prior information. The prior information is derived from a HR complex image obtained either by interpolating the LR image corresponding to the central LED or using previously reconstructed results during continuous imaging. To reduce the generation time and enhance the stability, we introduce physical constraints on the number of available LEDs for each illumination pattern during the update process, effectively reducing the solution space. Once the illumination mode is generated, we collect LR images by using it to illuminate and reconstruct the amplitude and phase of the sample through our proposed physical neural network. Thanks to its unsupervised nature, our model can adjust the acquisition time of LR images by reducing the number of illumination patterns as required during deployment. Generally, more illumination patterns with fewer LEDs per pattern help to obtain higher quality of the reconstruction result, while one containing fewer patterns with more LEDs per pattern helps to shorten the time of information acquisition and reconstruction processes. Features of our model for unsupervised training and adaptation effectively address the issue of dataset scarcity and enhance the model's generalization, rendering it more appropriate for practical applications.

This paper is organized as follows. Section 2.1 expounds upon the architecture and underlying principles of our proposed model. The specific methodology and computational procedures for generating the illumination mode are discussed in Section 2.2, while Section 2.3 presents the capture and reconstruction processes. The simulation and experiment principles are elucidated in the final segment of Section 2. In Section 3, simulations and experiments are conducted to validate the capability of our model to generate adaptive illumination modes with high reconstruction quality. Conclusions are subsequently synthesized in Section 4.

## 2. Method

Unlike the typical FPM system, we first generate a specific illumination mode which is appropriate for illuminating the sample based on the ideal imaging process. Then we employ this mode to illuminate the sample, capture the intensity of LR images, and reconstruct the target's amplitude and phase. We demonstrate our experimental system in Fig. 1.

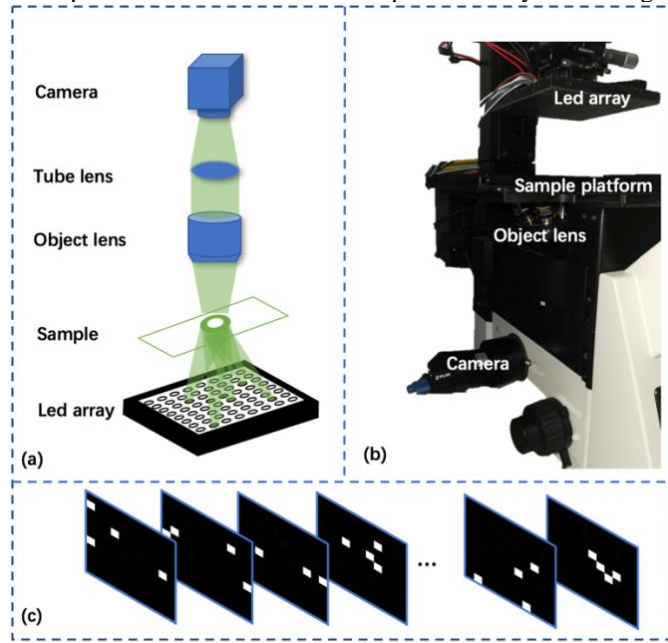

Fig. 1 (a) The theoretical model of FPM (b) The experimental system of our method (c) the generated illuminating model with our method

### 2.1 FPM principle and model structure

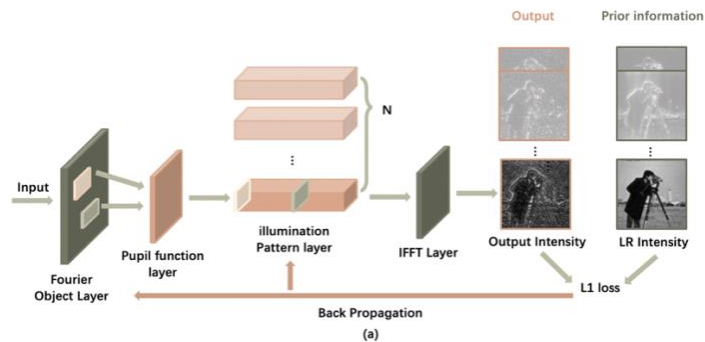

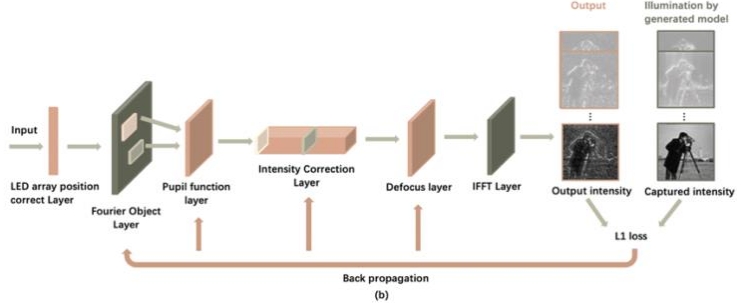

Fig. 2. (a)structure of the ideal process model CFPN for generating illumination mode.
(b)structure of CFPMN

In a forward imaging process, if the sample is much smaller than its distance to the LED array, the illumination light can be treated as a parallel plane wave. Each LR image corresponding to an individual LED offers information on different sub-spectrum areas of the sample, with the wave vector $(k_x, k_y)$ determined by the position of each LED and the distance between the LED array and the sample. In scenarios involving sequential illumination mode or position-multiplexed illumination mode, the imaging process can be regarded as performing FFT, low-pass filtering in the Fourier domain, iFFT, and intensity imaging in the spatial domain. It can be expressed as

$$I_n(x,y) = \left|\mathcal{F}^{-1}\{\mathcal{F}\{t(x,y)\} \cdot P(k_x, k_y)\}\right|^2 \quad (1)$$

The $t(x,y)$ denotes the sub-region on the frequency domain, where $(x,y)$ indicates the spatial location of the illuminating source. The Fourier transform is represented by $\mathcal{F}$, while its inverse process is represented by $\mathcal{F}^{-1}$. The pupil function of our system is denoted by $P(k_x, k_y)$, and $(k_x, k_y)$ maps the spatial position in the frequency domain. Finally, $I_n(x,y)$ represents the two-dimensional intensity of the final captured image.

Conventional FPM optimization algorithms require finding analytical differentiation, which is difficult when optimizing multiple parameters simultaneously. In contrast, FPMN [24] models the forward propagation process as an element-wise neural network, taking advantage of numerical differential methods to simplify the optimization process. FPMN's joint optimization capability and stability during training make it possible for future researchers to model and interpolate other parameters in the non-ideal state imaging process using a similar framework. Consequently, we establish a model sharing the similar framework with FPMN for our research and rewrite some layers using Pytorch's new complex data type to reduce complexity and enhance operational efficiency. In essence, our model can be viewed as an evolution of FPMN. Eq. (1) is rewritten accordingly.

$$I_n = \left|\mathcal{F}^{-1}\{\hat{o}(k - k_n) \cdot P(k_x, k_y)\}\right|^2, n = 1, 2, \ldots N \quad (2)$$

where $\hat{o}$ denotes the Fourier transform of the object function, $k_n$ represents the illumination vector and $P(k_x, k_y)$ have the same meaning as in Eq. (1). The variable $N$ represents the total number of LEDs that are present on the LED array.

We first rewrite the ideal Fourier Ptychography Neural Network (FPN) [24] as Complex Fourier Ptychography Neural Network (CFPN) and reconstructed to add our illumination pattern layer, which is used to generate the specific illumination mode, as shown in Fig. 12(a). The total number of illumination patterns in the mode is N. The two-original float-type sub-channels that represent the sample's complex function were merged into a single channel of complex type, which is the same for the inverse fast Fourier transform (IFFT) layer. For modeling the complex pupil function $P(k_x, k_y)$, we used the top ten Zernike coefficients, which is a classic two-dimensional phase distribution representation approach.

*2.2 Generate specific illumination mode*

During the process of generating a specific illumination mode, we use either the HR image from the last frame in the video dataset or the image obtained by bilinear interpolation from the LR image illuminated with the middle LED as the prior knowledge of the sample's phase and amplitude to simulate the imaging process. As each LED illuminates incoherently with the others, the final captured intensity in a multi-LED scenario can be regarded as a straightforward linear sum of the individual intensities produced by each LED. This can be expressed as

$$MI_p = \sum_{i=0}^{N} w_i \cdot I_i(x_i, y_i) \tag{3}$$

We use $MI_p$ to represent the ideal captured intensity, and $N$ represents the total number of LEDs. The $w_i$ parameter is used to correct for any fluctuations in the illumination intensity, which helps make our model more realistic. The $I_i(x_i, y_i)$ has the same meaning as in Eq. (1).

The process of obtaining the illumination mode is approached as a dynamic programming (DP) process. DP is a computational technique to solve optimization problems by breaking them down into smaller sub-problems and finding the optimal solution to each sub-problem. To tackle this challenge, a greedy algorithm was utilized, which is a type of algorithmic approach in computer science and optimization theory that makes locally optimal choices at each step in the hope of finding a global optimum. In other words, the best available option is chosen at each step without considering the overall effect on the final solution.

To start, the illumination mode is considered as a queue M for storing patterns, with the capacity ceiling of the queue set as the number of patterns contained in one mode. Initially, the queue is empty. To choose which LEDs to be turned on in a pattern, the extreme case where all LEDs on the available area are turned on is assumed, and the lighting weights for each LED are given the same initial value to ensure that all LEDs have an equal chance of being lit before a new pattern is fixed. After the intensity of the LR images is recorded, the network is tasked with reconstructing the exact sample information according to the collected images. This approach makes solving the inverse problem more difficult, but thanks to our modified CFPN, we can transform the original optimization problem into a loss function minimization problem, as illustrated in Fig 2(a). During the backpropagation process, the weights of LEDs that help reduce the loss function value will be increased significantly. We assign a weight of 1 to those LEDs that are clearly beneficial and set the rest to 0. This pattern is then fixed and added to the queue M. When M is empty, we run the propagation process with all LEDs on, fix and add the obtained pattern. When the queue M contains n (where n is not 0) patterns, we dequeuer all patterns in it to create an illumination mode containing n+1 patterns, and perform a network training process based on this mode. In the n+1th pattern, the LEDs not included in the first n patterns are initialized with the same weight, and the rest are set to 0. Similarly, we set the weights of the LEDs with high importance in the n+1th pattern to 1 and the weights of those with relatively low importance to 0 and add all n+1 patterns to the queue M. This algorithm is repeated until the queue is full. Consequently, for each observation, the illumination mode is specific to the sample and locally optimal.

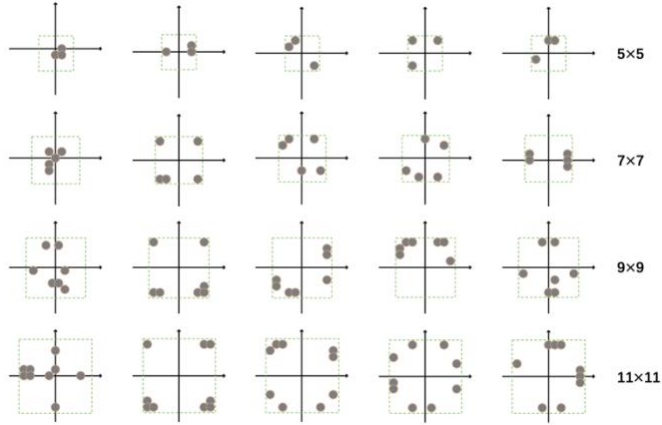

Fig. 3. An example illumination mode in which the yellow square represents the turned-on LEDs and another color means off (see **Visualization 1**).

Our model can also adaptively generate the illumination mode based on the number of different groups of patterns provided. However, the simulation process has a higher signal-to-noise ratio than the actual acquisition process. In particular, the dark field information is frequently damaged during actual capture. To enhance the practical application ability of our proposed model, we artificially divided the patterns into groups, with each group corresponding to a different size of the optimizable LED array area, as shown in Fig. 3. We have created four groups with optimized LED numbers of 25, 49, 81, and 121, each containing five patterns. It's worth mentioning that this division keeps the captured energy within a small fluctuation range under each pattern illumination, making it easier to adjust the exposure time during experimental acquisition process. The number of groups and the number of patterns within each group serve as adjustable hyper-parameters that can be modified as needed. Nonetheless, our model remains highly competitive even without grouping, as demonstrated in the follow-up experiment section.

Unlike the previous work of Delong Y et al. [24], we incorporate a normalization step during the forward propagation of the model. This approach improves the stability during the training process and is more consistent with the experimental LR image collection process, making the increase of LED brightness weight no longer a simple linear process. It also avoids the impact of different initialization values on the overall illumination mode. Moreover, it prevents the network from excessively increasing the brightness of some LEDs during the optimization process, which could cause the ignore of other LEDs that are also beneficial for minimizing the loss function value.

*2.3 Capture low-resolution images and reconstruct*

In this process, we illuminate the observed object with the mode generated and record the LR images. To enhance stability during training, we linear stretch the intensity values of the model output during forward propagation. This method is an extension of the approach used by Delong Yang et al. [24] and is described by Eq. (4).

$$MI_p = \frac{MI_p}{n * 16} \tag{4}$$

Where $MI_p$ represents the predicted captured intensity for the pth pattern, and $n$ represents the number of LEDs containing in the pattern. Rather than modeling an idealized process, we rewrite the FPMN network as the complex Fourier ptychography multi-parameter neural network (CFPMN), similar to the aforementioned CPFN, for a more accurate representation of the actual physical processes involved. This allows us to jointly optimize parameters representing other physical processes that may appear in the non-ideal imaging situation, while also inheriting and building upon the strong scalability of FPMN's framework. By following the work of Delong Yang el al. [24], we choose the L1-norm as the loss function in our approach, which is shown in the following equation.

$$loss = \frac{1}{n}\sum_{n=1}^{N}|I_n^{gt} - I_n^{predict}| \quad (5)$$

*2.4 Simulation and experiment principles*

In the simulation, we use ground truth to initialize the sample layer and simulate the propagation of light during imaging by performing a forward process of the CFPN mentioned above. The series of LR images output by the network were used to simulate the image intensity collected by the camera during the actual imaging process. In the reconstruction process, we use the LR image illuminated by the central LED as the initial value of the Fourier object layer. The simulation results can well reflect the difficulty of solving the inverse problems with different illumination mode. In the following experiments, we demonstrate that the reconstruction results of simulation and experiment are highly consistent. It is worth integrating that we adjust the exposure time of different patterns to ensure the capture of dark-field information. Undoubtedly, this adjustment may alter the overall brightness of the LR image, complicating the reconstruction process. To address this issue, we adjust the brightness of the LR images based on the approximate theoretical brightness obtained through simulation, as shown in the following equation.

$$E_{sn} = sum(F(CFPN, n)) \quad (6)$$
$$E_{en} = sum(I_n) \quad (7)$$
$$I_n = \frac{E_{sn}}{E_{en}} \cdot I_n, (n = 1 \dots N) \quad (8)$$

The energy of the simulation output of the LR image is represented by $E_{sn}$, while $F$ represents the forward process of our model. Here, $I$ and $N$ have the same meaning as mentioned previously.

## 3. Simulations and experiments

*3.1 Experimental results of USAF resolution target with our model*

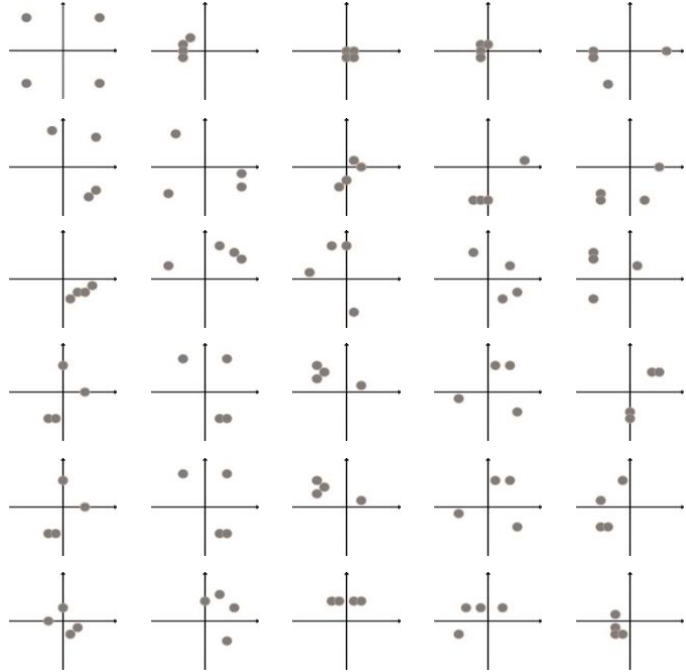

Fig. 4 The illuminating model of 11×11 available LEDs generated by our model.

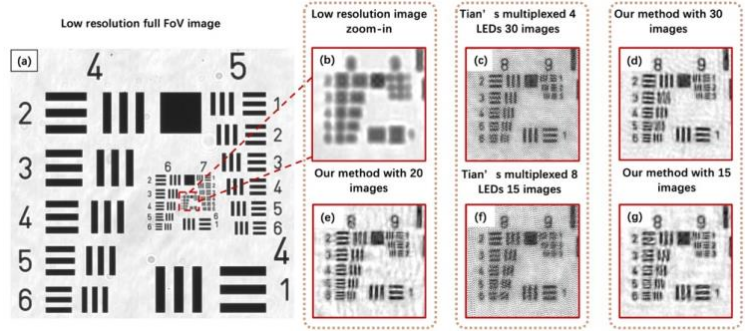

Fig. 5. Experiment results for multiplexed illumination of a resolution target. (a) The captured image illuminated by the middle LED. (b) The low-resolution area without reconstructing. (c) The reconstruction result from 30 images illuminated by multiplexed 4 LEDs which is heuristic design. (d) The reconstruction result from 30 images illuminated by our generated illumination mode. (e) The reconstruction result from 20 images illuminated by our generated illumination mode. (f) The reconstruction result from 15 images illuminated by multiplexed 8 LEDs which is heuristic design. (g) The reconstruction result from 15 images illuminated by our generated illumination mode.

In this section, we demonstrate the effectiveness and validity of our model by choosing the method of Tian's heuristic designed multiplexed 4 LEDs [4,14,28] as the benchmark and extending it to the situation of 8 LEDs. For the experiment, we used a 4X objective with an NA of 0.4542 and a LED array containing 11×11 LEDs, with a distance of 5mm between each of them, placed 100mm beneath the sample. By following the work of Delong Yang et al. [24], we limit the captured images to 256×256 pixels while the HR result was reconstructed to 1024×1024 pixels. To generate an illumination mode with 11x11 available LEDs, containing 30 patterns and 4 LEDs were turned on in each pattern, we used our model as shown in Fig. 4. Interestingly, our model tends to include only bright-field LEDs or dark-field LEDs in one pattern, unlike random methods. This occurs because bright and dark field LEDs offer distinct information, and our model develops this differentiation capability through learning. The quality of the reconstructed result degrades as the number of captured LR images decreases, which is also observed in previous work [4,14,29], and we have the same phenomenon in our simulations. We attribute this to the reduction in the average spatial distance between two adjacent lit LEDs when decreasing the total number of available LEDs. Turning on too many LEDs in one pattern increases the difficulty of reconstruction. To make the training process of the network more stable, we used the LR image obtained by linear interpolation of the image illuminated with the center LED or prior HR image to initialize the sample layer. In other words, we additionally collected a LR image as the initial value, which is also used to generate the specific illumination mode when lacking other prior information. The experimental results illuminated by different numbers of patterns are shown in Fig. 5.

*3.2 Simulation results of USAF resolution target with our model*

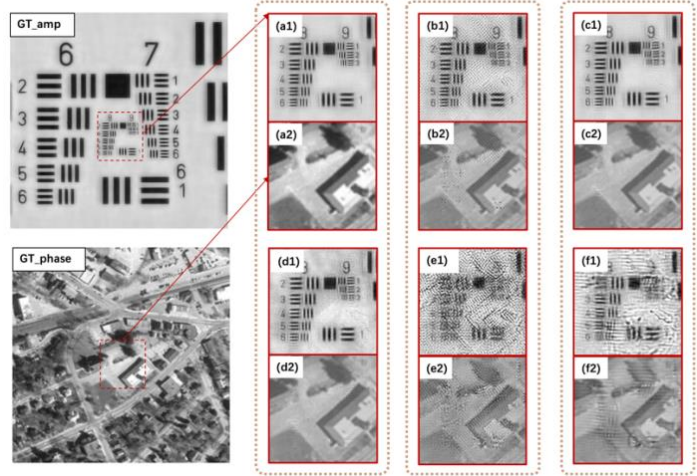

Fig. 6. Simulation results for multiplexed illumination of a hypothetical sample. (a1) The ground truth for the amplitude within the central 256x256 area. (a2) The ground truth for the phase within the central 256x256 area. (b*) The reconstruction result from 30 images illuminated by Tian's multiplexing 4 LEDs. (c*) The reconstruction result from 20 images illuminated by our generated model. (d*) The reconstruction result from 20 illuminated by our illuminating model generated depending on incoherent information. (e*) The reconstruction result from 15 images illuminated by Tian's multiplexing 8 LEDs. (f*) The reconstruction result from 15 images illuminated generated by our model.

**Table 1. Comparison of the performance with different methods and illuminating model size**

| Method | Evaluation metric(amplitude) | | | | |
|---|---|---|---|---|---|
| | L1 | SSIM | PSNR | NIQE | LPIPS |
| 30 images (Tian's multiplexing 4 LEDs [4,14,21]) | 240 | 0.495 | 24.3 | 21.0 | 0.220 |
| 20 images (our method based on incoherent information) | 197 | 0.704 | 25.2 | 20.3 | 0.181 |
| **20 images (our method)** | **48.4** | **0.842** | **31.3** | **19.0** | **0.085** |
| 15 images (Tian's multiplexing 8 LEDs [4,14,21]) | 1730 | 0.176 | 15.8 | 21.9 | 0.349 |
| 15 images (our method) | 768 | 0.351 | 19.3 | 22.7 | 0.308 |

The high-level agreement between the simulations and experiments demonstrates the effectiveness of our simulation method, as shown in Fig. 5 and 6. Moreover, compared to actual experiments, the ground truth of the simulation experiments is more accurate and easier to obtain. Thus, we use the performance of the simulation results to evaluate the quality of different illumination modes in subsequent experiments. When observing samples with complex phases, our generated mode shows significant advantages in terms of acquisition effectiveness and optimization effects, which are difficult to match with other methods. As the number of captured images decreases, the reconstruction result also declines, as shown in Fig. 6. However, we can still make a trade-off between reconstruction speed and clarity based on different application scenarios since our reconstruction process remains stable under different pattern numbers. Additionally, we use different illumination modes to demonstrate the specificity of our method, as shown in Table 1. We employ the metrics of SSIM, PSNR, NIQE, and LPIPS [29-32], which are commonly used by super-resolution works in the field of computer vision, to evaluate the effectiveness of our method. It is worth mentioning that the illumination mode based on incoherent information has almost the same performance as the Tian's multiplexed method, which highly proves that our method is adaptive to the specific samples.

3.3 The experiments of the biological sample

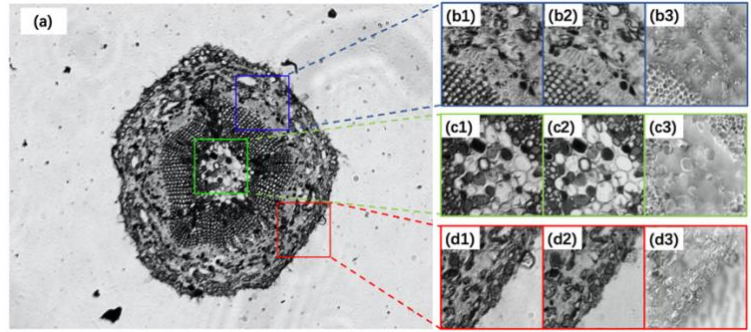

Fig. 7 The experimental results of plant stem cross-section illuminated with our illumination mode. (a) The intensity of full image illuminated by center LED. (*1) The captured low-resolution image of center LED. (*2) The reconstructed amplitude of different regions. (*3) The reconstructed phase of different regions.

Unlike the resolution target, the phase and spectrum characteristics of actual biological samples tend to be more complex. To demonstrate the efficiency of our model in practical applications, we utilized plant stem cross-section as our observation sample. We kept the same parameters mentioned in Section 3.1 and captured a series of LR images with a size of 1536×2048. Then we selected 256×256 sub-images from different regions and individually reconstructed their phase and amplitude. The results, which are shown in Figure 7, provide conclusive evidence of the effectiveness of our proposed method.

## 4. Conclusion and future work

In this paper, we have presented an adaptive coded illumination technique for FPM using an extended physical neural network model. Our method can provide adaptive illumination modes specific to different samples, and can further compress LR image acquisition time based on unsupervised optimization, significantly expanding the application scope of the FPM system. Unlike the data-driven method, our method is established according to physical rules, enhancing the model interpretability. The parameters of the illumination pattern layer are optimized simultaneously with those of the sample layer to achieve adaptive generation. As every LED activated in our illumination mode exhibits the same brightness, this significantly alleviates the strain on the hardware. Our method has achieved state-of-the-art results with great advantages in high-frequency information recovery. Additionally, we observed that the initial value selection for the physical neural network framework can have an impact on the final optimization result. Thus, we plan to dedicate more research to optimization algorithms, including exploring a general update of the network that differs from the original FPM method in our future work.

**Funding.** National Key Research and Development Program of China (No. 2021YFC2202404); National Natural Science Foundation of China (62275020).

**Disclosures.** The authors declare no conflicts of interest.

**Data availability.** Data underlying the results presented in this paper are not publicly available at this time but may be obtained from the authors upon reasonable request.